\documentclass{ws-ijmpa}
\usepackage[super,compress]{cite}
\usepackage{graphicx}
\begin{document}
\markboth{Mathews et al.}{Constraints on the Birth of the Universe}

\catchline{}{}{}{}{}
%
%%%%%%%%%%%%%%%%%%%%%%%%%%%%%%%%%%%%%%%%%%%%%%%%%%%%%%%%%%%%%%%%%%%%

\title{Constraints on the birth of the universe and  origin of cosmic dark flow}

\author{G. J. Mathews, M. R. Gangopadhyay, P. Garnavich, B. Rose}

\address{Center for Astrophysics, Department of Physics, University of Notre Dame\\ Notre Dame, IN 46554, USA\\
gmathews@nd.edu, mgangopa@nd.edu, pgarnavi@nd.edu, brose3@nd.edu
}
\author{K. Ichiki}

\address{Department of Physics, Nagoya University\\
 Nagoya 464-8602, Japan
%ichiki.kiyotomo@c.mbox.nagoya-u.ac.jp 
}
\author{T. Kajino}

\address{National Astronomical Observatory, 2-21-1, Osawa\\
Mitaka, Tokyo 181-8588, Japan\\
and 
University of Tokyo, Department of Astronomy
7-3-1 Hongo, Bunkyo-ku\\
 Tokyo 113-0033, Japan }

\author{Dai Yamazaki}

\address{University Education Center, Ibaraki University\\
 2-1-1, Bunkyo, Mito\\
Ibaraki 310-8512, Japan
%dai@mx.ibaraki.ac.jp
 }

\maketitle

\begin{history}
\received{Day Month Year}
\revised{Day Month Year}
\end{history}

\begin{abstract}
We summarize three recent efforts to constrain the first few moments of cosmic creation before and during the epoch of inflation.  We consider two means to explain a slight dip in the power spectrum of the cosmic microwave background for multipoles in the range of $\ell= 10-30$ from both the {\it Planck} and {\it WMAP} data.  We show that such a dip could be the result of resonant creation of a massive particle that couples
 to the inflaton field.   For 
best-fit models, the epoch of resonant particle creation reenters the
horizon at wave numbers of $k_* \sim 0.00011 \pm 0.0004 $  ($h$
Mpc$^{-1}$).  The amplitude and location of these features correspond
to the creation of a number of degenerate fermion species of mass $\sim 15/\lambda^{3/2} $ $m_{pl}$ during
inflation where $\lambda$ is the coupling constant between the inflaton field and the
created fermion species.  Alternatively, one can explain the existence of such a dip as due to a jump in the
inflation generating potential.  We show that such a jump can also resolve the excessively large dark flow predicted from the M-theory landscape.  Finally, we summarize our efforts to quantify  constraints on the  cosmic dark flow from a new analysis of the Type Ia supernova distance-redshift relation.

\end{abstract}

\ccode{PACS numbers: 98.80.Cq, 98.80.Es, 98.70.Vc}

%\tableofcontents

\section{Introduction}	

Analysis\cite{PlanckXIII,PlanckXX} of the power spectrum of fluctuations in the 
cosmic
microwave background (CMB) 
 provides powerful constraints on the physics of
the very early universe. 
The most popular account for the
origin of  the primordial power spectrum is based upon quantum fluctuations 
generated
during the inflationary epoch\cite{Liddle,cmbinflate}.  
Subsequently, acoustic oscillations of the photon-baryon fluid distort
this to produce the observed features in the angular power spectrum of
temperature fluctuations in the CMB and the spatial power spectrum of
matter density fluctuations.

In this context, 
there now  exists the highest resolution yet available in the  determination
of the power spectrum of the CMB  from the {\it Planck} Satellite \cite{PlanckXIII,PlanckXX}.  
In Mathews et al.\cite{Mathews15} we made note of a peculiar feature visible in the power spectrum 
near multipoles $\ell = 10-30$.  This is an interesting region in the CMB power spectrum because it corresponds to 
angular scales that are not yet in causal contact, so that the observed power spectrum is close to  the true primordial power spectrum.  

\begin{figure}[htb]
\includegraphics[width=3.5in,clip]{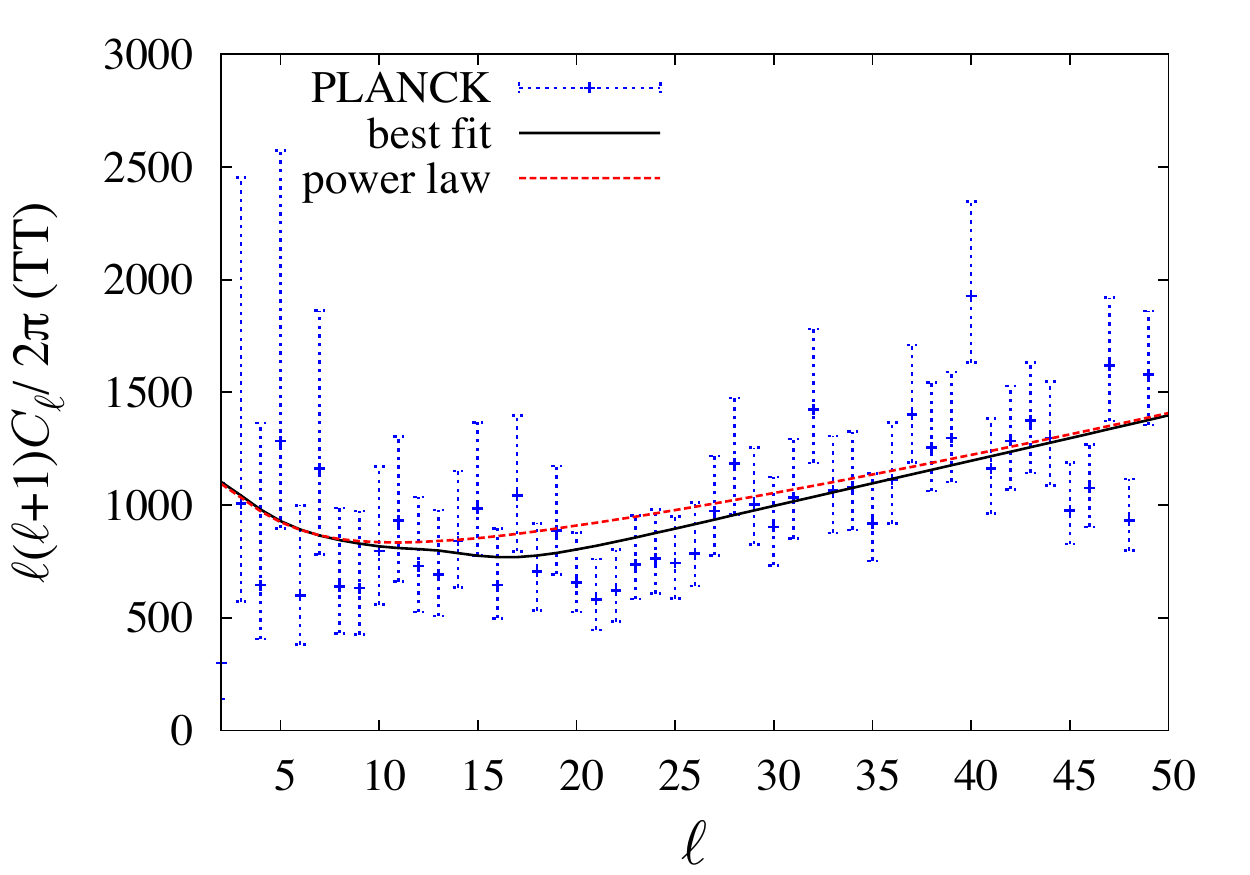} 
\caption{CMB power spectrum in the range of $\ell = 0-50$  from Mathews et al.\cite{Mathews15} Points with error bars are from the 
{\it Planck} Data Release \cite{PlanckXIII}.  The dashed line shows the best standard $\Lambda$CDM fit to the {\it Planck} CMB power spectrum
based upon a power-law primordial power spectrum.  The solid line shows the best fit for a model with resonant particle creation during inflation.}
\label{fig:1}
\end{figure}

An illustration of the {\it Planck} observed power spectrum in this region is visible on Figure \ref{fig:1} from Mathews et al.\cite{Mathews15}.
Although the error bars are large, there is a noticeable systematic deviation  in the range $\ell = 10-30$ below the  best fit based upon the standard $\Lambda$CDM cosmology with a power-law primordial power spectrum.  
This same feature is visible in the CMB power spectrum from the Wilkinson Microwave Anisotropy Probe ({\it WMAP}) \cite{WMAP9}, and hence, is likely a true feature in the CMB power spectrum.  This can be interpreted as an artifact of the cosmic variance \cite{PlanckXIII}, or a phase transition in the inflation-generating potential \cite{wiggleV, Gangopadhyay15, Iqbal15}.  However, here 
we adopt the the premise that this could be a real feature in the primordial power spectrum produced by new trans-Planckian physics occurring  near the end of the
inflation epoch.  In particular, we show that  this feature is well represented by the resonant creation \cite{chung00,Mathews04} of Planck-scale particles that
 couple to the inflaton field as shown  by the solid line in Figure \ref{fig:1} and we now describe in detail.

This interpretation has the
intriguing aspect that, if correct, an opportunity emerges to use the
CMB  as a probe of physics at the  Planck scale ($m_{pl} \sim
10^{19}$ GeV).
Indeed, massive particles generically exist in Planck
scale compactification schemes of string theory from the Kaluza-Klein states,
winding modes, and the massive (excited) string modes.
%Hence, the existence of Planck-scale
%mass particles which couple to the inflaton is a reasonable scenario.  
%What is, perhaps,  not as likely  in the paradigm considered here  is
We have explored\cite{Mathews15} the possibility that the mass of some Planck-scale particles lies in the range of the vacuum energy associated with 
 the last 9-10  $e$-folds of  inflaton accessible to observation. 
%Nevertheless, in view of the importance of such a discovery, it is worthwhile to examine this possibility based upon the
%currently available CMB power spectra.

\section{Inflation Resonant Particle Production}

The details of the  resonant particle creation paradigm have been explained in Refs.~\citen{chung00,Mathews04, Mathews15}.
Here, we summarize essential features.
 In the basic inflationary picture, a rapid early expansion of the
universe is achieved through the vacuum energy from an inflaton field.
In the minimal extension from the basic picture considered here, the
inflaton is postulated to couple to massive particles
whose mass is of order the inflaton field value.  These particles
are then resonantly produced as the field obtains a critical value
during inflation.  If even a small fraction of the 
inflaton field is affected  in this way, it can produce an observable feature in
the primordial power spectrum. In particular, there can  be either and excess\cite{chung00,Mathews04},
or a dip\cite{Mathews15} in the in the power spectrum. Such a dip, will
occur at an  angular scale corresponding to when the
epoch of resonant particle creation crossed the Hubble radius during inflation.

We note that  particle creation corresponding to an imaginary
part of the effective action of quantum fields has been considered in \cite{Starobinsky02}.  In that case the same creation should occur at the present time.
In that case compatibility with
the diffuse $\gamma$-ray background can be used to rule out the possibility of measurable effects from this type of
trans-Plankian particle creation in the CMB anisotropy.
However, the effect of interest here is a perturbation in the simple scalar field due to direct coupling to Plack-mass particles at energies for which the inflation potential is 
comparable to the particle mass and cannot occur at the present time.  The present scenario, therefore is  not
constrained by the diffuse gamma-ray background.

In the simplest slow roll approximation \cite{Liddle,cmbinflate} for the
generation of density perturbations during inflation, the amplitude,
$\delta_H(k)$, of a density fluctuation when it crosses the Hubble
radius is just,
\begin{equation}
\delta_H(k) \approx {H^2
\over 5 \pi \dot \phi}~~,
\label{pert}
\end{equation}
where $H$ is the expansion rate, and $\dot \phi$ is the rate of change of
the inflaton field when the comoving wave number $k$ crosses the
Hubble radius during inflation.  If resonant particle production
affects the inflaton field, then the conjugate momentum in
the field $\dot \phi$ is altered.  This causes either an increase or a diminution in
$\delta_H(k)$ (the primordial power spectrum) for those wave numbers which
exit the horizon during the resonant particle production epoch.

%Of course when $\dot{\phi}$ is changing due to particle production,
%the acceleration  $\ddot{\phi}$ may not be negligible.  This would result  in a correction to
%Eq.~(\ref{pert}).  However, in \cite{chung00}, this correction was analyzed
%and found to be $<<20 \%$.  Hence, for our purposes we can ignore this correction.

For the application here, we adopt a  positive Yukawa coupling of strength $\lambda$ between the inflaton field $\phi$ and the field $\psi$ of
N fermion species.   This differs from \cite{chung00, Mathews04} who adopted a negative Yukawa coupling.
With our choice, the total lagrangian density including the inflaton scalar field $\phi$, the dirac fermion field, and
the Yukawa coupling term is then written,
\begin{eqnarray}
{\cal L}_{\rm tot} &=& \frac{1}{2}\partial_\mu \phi \partial^\mu \phi - V(\phi) \nonumber \\
&+& i \bar \psi \gamma^\mu \psi - m \bar \psi  \psi + N \lambda \phi \bar \psi  \psi ~~.
\end{eqnarray}
For this Lagrangian, it is obvious that the fermions have an effective mass of
\begin{equation} 
M(\phi) = m - N \lambda \phi~~.
\end{equation}
This  vanishes
for a critical value of the inflaton field,
$\phi_* = m/N \lambda$.  Resonant fermion production
will then occur in a narrow range of inflaton field amplitude
around $\phi = \phi_*$.

As in Refs.~\refcite{Mathews15,chung00,Mathews04} we label the epoch at which particles are created
by an asterisk.  So the cosmic scale factor is labeled $a_*$ at the
time $t_*$ at which resonant particle production occurs.  Considering
a small interval around this epoch, one can treat $H = H_*$ as
approximately constant (slow roll inflation).  The number density $n$
of particles can be taken as zero before $t_*$ and afterwards as $n =
n_*[a_*/a(t)]^{3}$.  The fermion vacuum expectation value can then be
written,
\begin{equation}
 \langle \bar \psi \psi \rangle = n_* \Theta (t-t_*) \exp{[-3 H_*(t-t_*)]} ~~.
 \end{equation}
where $\Theta$ is a step function.

Then following the previous derivations\cite{chung00,Mathews04,Mathews15}, we have the following modified equation of motion for the scalar field coupled to $\psi$:
\begin{equation}
\ddot \phi + 3 H \dot \phi = -V'(\phi) +  N \lambda \langle \bar \psi \psi \rangle ~~,
\end{equation}
where $V'(\phi) = dV/{d\phi}$.
In the slow roll approximation, ($\ddot \phi \sim 0,~H\sim H_*,~V'(\phi) \sim V'(\phi_*$) 
  this differential equation after particle creation $(t>t_*)$ is then the same as in \cite{chung00,Mathews04} but with a sign change for the coupling term, i.e.
\begin{eqnarray}
\dot \phi(t > t_*) &=& \dot \phi_* \exp{[-3H(t-t_*)]}\nonumber \\
& -& \frac{V'(\phi)_*}{3 H_*} \bigl[ 1 - \exp{[-3H(t-t_*)]}\bigr] \nonumber \\
&+&  \frac{N \lambda }{3 H_*}n_* (t-t_*) \exp{[-3 H_*(t-t_*)]} ~~.
\end{eqnarray}
The physical interpretation here is that the rate of change of the scalar field rapidly increases due to the coupling to particles created at the resonance 
$\phi = \phi_*$.  

Then, using Eq.~(\ref{pert})
for the fluctuation as it exits the horizon, and  constant $H \approx H_*$ in the slow-roll condition
along with
\begin{equation}
d\ln{a} = Hdt ~~,
\end{equation}
 one obtains  the perturbation in the primordial power spectrum as it exits the horizon:
\begin{equation}
\delta_H = \frac{[\delta_H(a)]_{N \lambda = 0}}{1 + \Theta (a - a_*)( N \lambda n_*/\vert \dot \phi_*\vert H_*) (a_*/a)^3 \ln{(a/a_*)}} ~~.
\label{deltahnew}
\end{equation}
Here,  it is clear that the power in the fluctuation of the inflaton field will diminish as the particles are resonantly created when the universe
grows to some critical scale factor $a_*$.

Using $k_*/k = a_*/a$,
the perturbation spectrum Eq.~(\ref{deltahnew})
can be reduced \cite{Mathews04, Mathews15} to a simple 
two-parameter function.
\begin{equation}
\delta_H (k) = \frac{[\delta_H(a)]_{N \lambda = 0}}{1 + \Theta (k-k_*)A (k_*/k)^3 \ln{(k/k_*)}} ~~.
\label{perturb}
\end{equation}
where the amplitude  $A$ and characteristic wave number $k_*$ ($k/k_*
\ge 1$) can be fit to the observed power spectra from the relation:
\begin{equation}
k_* = \frac{ \ell_* }{ r_{lss}}~~, 
\end{equation}  
where $r_{lss} $ is the
comoving distance to the last scattering surface, taken here to be 14 Gpc.

%We have MCMC fit the Planck CMB power spectrum using the perturbation of Eq.~(\ref{perturb}) 
%plus a a standard power-law $k^n$ primordial power spectrum with other cosmological parameters fixed at the maximum likelihood values from the Planck analysis..
 The connection between 
resonant particle creation and the CMB temperature fluctuations 
is straightforward.  As usual, temperature fluctuations 
are expanded in spherical harmonics, $ \delta T/T =\sum_l\sum_m
a_{lm}Y_{lm}(\theta,\phi)$ ($2 \le l<\infty$ and $-l \le m \le l$). 
The anisotropies are then described by the angular 
power spectrum, $C_l= \langle |a_{lm}|^2\rangle$, as 
a function of multipole number $l$.  One then merely
requires the conversion from perturbation spectrum $\delta_H (k)$
to angular power spectrum $C_l$.  This is easily 
accomplished using the {\it CAMB} code \cite{Camb}.
When converting to the angular power spectrum,
the amplitude of the narrow particle creation
feature  in $\delta_H(k)$ is spread over many values of $\ell$.
Hence, the particle creation feature looks like a broad dip in the power spectrum.

We have made a  multi-dimensional 
 Markov Chain Monte-Carlo
analysis \cite{Christensen,Lewis} of the 
CMB using the {\it Planck}  data \cite{PlanckXIII} and the {\it CosmoMC} code \cite{Lewis}.   
For simplicity and speed in the present study we
only marginalized over 
parameters which do not alter the matter or CMB transfer functions. Hence, we only varied $A$ and $k_*$, along with the six parameters, 
 $\Omega_b h^2, \Omega_c h^2, \theta, \tau, n_s, A_s$.  Here,  $\Omega_b $ is the baryon content, $ \Omega_c $ is the cold dark matter content,  $\theta$ is the acoustic peak angular scale, $\tau$ is the optical depth, $n_s$ is the power-law spectral index, and $ A_s$ is the normalization.
%Hence, the set of free parameters in the analysis is ($n_s, A_s, \log{(k_*)},
%A,_* $), where $n_s$ is the spectral index, 
%$A_s$ is the overall amplitude of the primordial power spectrum,
As usual,
both $n_s$ and $A_s$ are normalized 
at $k = 0.05$ Mpc$^{-1}$. 

\begin{figure}
\includegraphics[width=3.5in,clip]{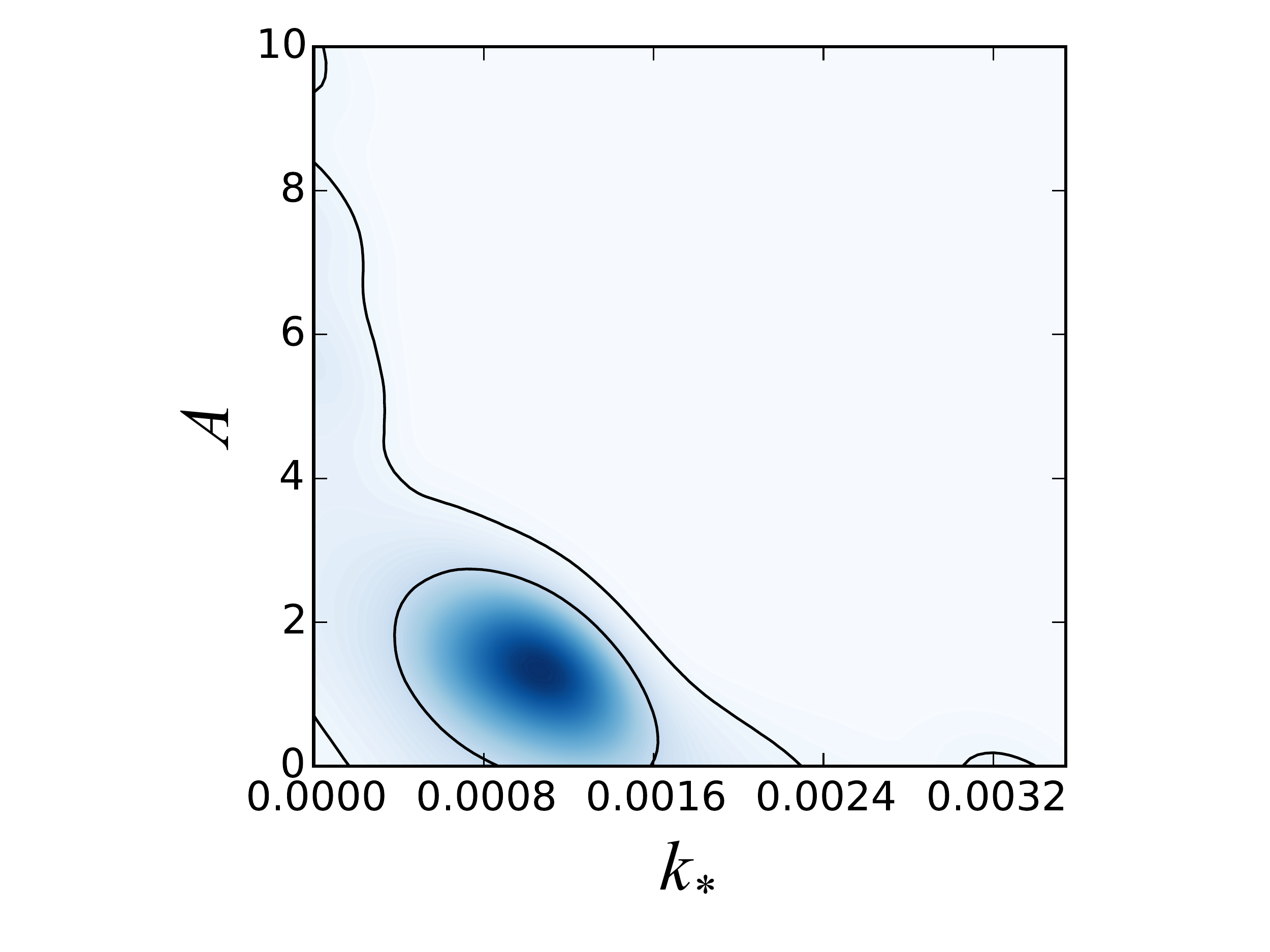} 
\caption{Constraints on parameters $A$ and $k_*$ from the
MCMC analysis\cite{Mathews15} of the CMB power spectrum.
Contours show 1 and  2$ \sigma$ limits.  The horizontal
axis is in units of ($h$ Mpc$^{-1}$). }
\label{fig:2}
\end{figure}

Figure \ref{fig:2} shows contours of likelihood for the resonant particle creation parameters,
$A$ and $k_*$.  Adding this perturbation to the primordial power spectrum improves the total 
$\chi^2$ for the fit from 9803 to 9798.  One expects that the effect of interest here would only make a small change ($\Delta \chi^2 = 5$) in the overall fit because  
it only affects a limited range of $l$ values.  Nevertheless, from
 the likelihood contours we can deduce mean value of $A = 1.7 \pm 1.5$ with a maximum likelihood value of   $A = 1.5$,  and a mean value of $k_* = 0.0011\pm 0.0004 ~h~{\rm Mpc}^{-1}$

The values of $A$ and $k_*$ determined from from the CMB power spectrum
relate to the inflaton coupling $\lambda$ and fermion mass $m$, for a
given inflation model via Eqs.~(\ref{deltahnew}) and (\ref{perturb}). 
\begin{equation}
A  = |\dot{\phi}_*|^{-1} N \lambda
n_* H_*^{-1} 
\end{equation}

The coefficient $A$
can be related directly to the coupling constant $\lambda$ 
using the approximation
\cite{chung00,birrellanddavies,Kofman:1997yn,Chung:1998bt} for the
particle production Bogoliubov coefficient 
\begin{equation}
|\beta_k|^2 = \exp\left( \frac{-\pi k^2}{a_*^2 N \lambda |\dot \phi_*| }\right).
\end{equation}

Then,
\begin{equation}
\label{eq:nstar}
n_* = \frac{2}{\pi^2}\int_0^\infty dk_p \, k_p^2 \, 
|\beta_k|^2 =
\frac{N \lambda^{3/2}}{2\pi^3}
|\dot{\phi}_*|^{3/2}~~ .
\end{equation}
This gives
\begin{eqnarray}
A & = & \frac{ N \lambda^{5/2}}{2 \pi^3}
\frac{\sqrt{|\dot{\phi}_*|}}{H_*}\\
& \approx & \frac{N \lambda^{5/2}}{2 \sqrt{5} \pi^{7/2}}
\frac{1}{\sqrt{\delta_H(k_*)|_{\lambda=0}}} ~~,
\label{eq:alamrelation}
\end{eqnarray}
where we have used the usual approximation for the primordial slow
roll inflationary spectrum \cite{Liddle,cmbinflate}.  This means that
regardless of the exact nature of the inflationary scenario, for any
fixed inflationary spectrum $\delta_H(k)|_{\lambda=0}$ without the
back reaction, we have the particle production giving  a dip of the
form Eq.~(\ref{perturb}) with the parameter $A$ expressed in terms of
the coupling constant through Eq.~(\ref{eq:alamrelation}).  Given that
the CMB normalization requires $\delta_H(k)|_{\lambda=0}\sim
10^{-5}$, we then have 
\begin{equation}
A \sim 1.3 N  \lambda^{5/2}.
\end{equation}
Hence, for the maximum likelihood value of $A \sim 1.5 $, we have 
\begin{equation}
N \approx \frac{(1.1^{+1.1}_{-0.9})}{\lambda^{5/2}} \sim \frac{1}{\lambda^{5/2}}~~.  
\label{Nconst}
\end{equation}
 So,   $\lambda \le 1$ requires $N>1$ as expected.

The fermion particle mass $m$ can then be deduced from $m = N \lambda \phi_*$.  From Eq.~(\ref{Nconst}) then we have
$m \approx   \phi_*/\lambda^{3/2}$.
For this purpose, however, one must adopt a specific form for the inflaton potential to determine $\phi_*$ appropriate to the scale $k_*$.
Here we adopt a general monomial potential whereby:
\begin{equation}
V(\phi) = \Lambda_\phi m_{pl}^4 \biggl(\frac{\phi}{m_{pl}}\biggr)^\alpha~~,
\label{Vphi}
\end{equation}
for which  there is a simple  analytic relation \cite{Liddle} 
between  the value of $\phi_*$ and the number of e-folds ${\cal N}(k_*)$ between when $k_*$ exits the horizon and the end of inflation, i.e.
\begin{equation}
{\cal N}(k_*) = \frac{1}{m{\rm pl}^2} \int_{\phi_{end}}^{\phi_*} \frac{V(\phi)}{V'(\phi)} d\phi~~,
\end{equation}
implies
\begin{equation} 
\phi_* = \sqrt{2 \alpha {\cal N}} m_{pl}~~.
\end{equation}

For $k_* = 0.0011\pm 0.0004 ~h~{\rm Mpc}^{-1}$, and $k_H = a_0 H_0 = (h/3000)$ Mpc$^{-1} \sim 0.0002,$ we have ${\cal N} - {\cal N}_*  = ln{(k_H/k_* )} <1$.  Typically one expects  ${\cal N}(k_*) \sim  {\cal N} \sim 50$  although one can have the number of e-folds as low as ${\cal N} \sim 25$ in the case of thermal inflation \cite{Liddle}.  So, for a monomial potential with   $\alpha = 2$ we have $\phi_* \sim 10- 14~ m_{pl}$.  On the other hand, the limits on the tensor to scalar ration from the {\it Planck} analysis \cite{PlanckXX} are more consistent with  $\alpha = 1$ ($\phi_* = 7-10 ~m_{pl}$), or even $\alpha = 2/3$ ($\phi_* = 6-8 ~m_{pl}$).  
Hence,  we have roughly the constraint,  
\begin{equation}
m \sim  6-14  ~\frac{m_{\rm pl}}{\lambda^{3/2}}~~, 
\end{equation}
well in excess of the Planck mass and independently of the number of degenerate species.

As noted above, it is natural\cite{chung00} to have such trans-Plankian massive particle in numerous extensions of the Standard Model
such as supergravity and superstring theories.  Such extra-dimensional theories generally  contain
a spectrum of particles with masses well in excess of the Planck mass. The extra-dimensions
are compact and smaller than the three large spatial dimensions. It is, therefore, possible to dimensionally reduce the
system to  obtain an effective (3+1) dimensional theory that produces a tower of Kaluza-Klein (KK) states \cite{Kolb84,Lewis03}.
The  mass of states in this tower is  of order of the inverse size of the extra dimension. Since the extra dimensions are expected to
have a size characteristic of the Planck length, these KK states therefore have masses in excess of the Planck masses. Furthermore, these KK
states can be  nearly degenerate, with the level of degeneracy depending upon the geometrical structure of the compact space.
It is, therefore, natural to deduce  that  a large number of nearly degenerate fermions existed during inflation with a mass well in excess of  $m_{pl}$,
and that these particles couple to the inflaton field that drives inflation.  For our purposes  this degeneracy factor is described 
by the parameter $N$, and in such theories   $N$ can easily of the order of 100.  Hence, such a detectable
signature in the CMB power is not unexpected.

\section{Matter Power Spectrum }

It is straight forward to determine the matter power spectrum.  To
convert the amplitude of the perturbation as each wave number $k$ enters
the horizon, $\delta_H(k)$, to the present-day power spectrum, $P(k)$,
which describes the amplitude of the fluctuation at a fixed time, one
must make use of a transfer function, $T(k)$ \cite{efstathiou} which
is easily computed using the {\it CAMB} code  \cite{Camb} for various
sets of cosmological parameters (e.g.~$\Omega$, $H_0$, $\Lambda$,
$\Omega_B$).  An adequate approximate expression for the structure power
spectrum is
\begin{equation}
\frac{k^3}{2\pi^2}P(k) = \left( \frac{k}{aH_0} \right)^4 T^2(k)
\delta^2_H(k) \ .
\end{equation}
This expression is only valid in the linear regime,
which in comoving wave number is up to approximately $k ^<_\sim
0.2~h$ Mpc$^{-1}$ and therefore
adequate for our purposes.
However, we also correct for
the nonlinear evolution of the power spectrum \cite{Peacock}.

\begin{figure}
\includegraphics[width=3.5in,clip]{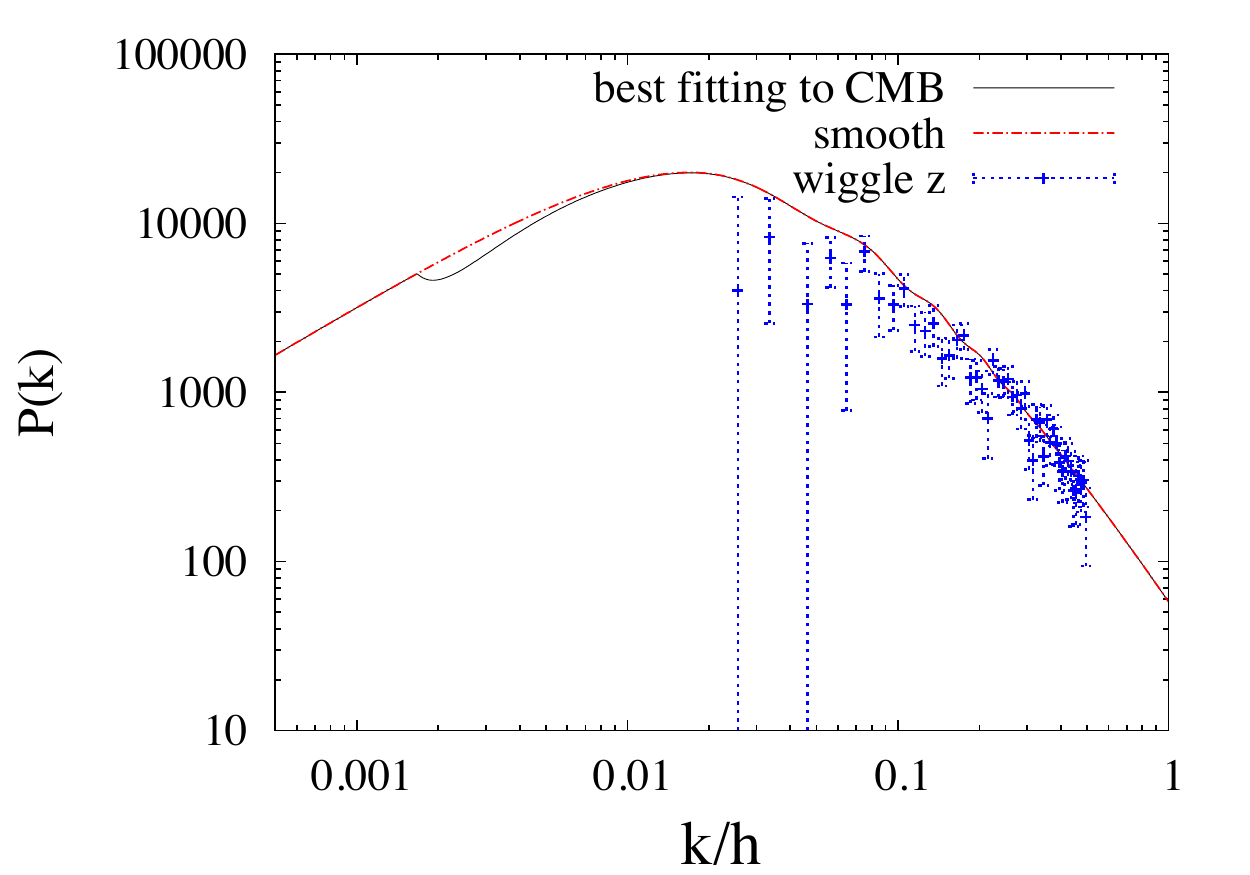}
 \caption{Comparison of the observed galaxy cluster function
from  \cite{wiggle}
 with the spectrum implied from
the fits to the matter power spectrum with (solid line)  and without (dashed line)
resonant particle creation during inflation from Ref.~\citen{Mathews15}.}
\label{fig:3}
\end{figure}

Figure \ref{fig:3} shows the  matter power spectrum from the {\it Wiggle-Z Dark Energy Survey} \cite{wiggle} compared to the computed
maximum likelihood power spectrum with and without
the perturbation due to the resonant particle creation.
Unfortunately, the  perturbation is on a scale too large to be probed by the observed matter power spectrum.

\section{Low Velocity Cosmic Dark Flow and the CMB power spectrum} 
An alternative way to explain the dip for $\ell = 10-30$ has been pointed out by a number of authors\cite{PlanckXX, wiggleV} as due to an abrupt jump in the inflation-generating potential.  In normal slow roll inflation, 
\begin{equation}
\dot \phi = \frac{dV(\phi)/d\phi }{3 H}~~,
\end{equation}
Hence, a large value for $dV(\phi)/d\phi$ can also produce a jump in the primordial power spectrim of Eq.~\ref{pert}.
In Gangopadhyay and Mathews\cite{Gangopadhyay15} we have explored this possibility with a potential of the form 
\begin{equation}
 V(\varphi ) = \frac{1}{2}m\varphi ^{2} + \lambda M _{Pl}^{4} \exp{[(-\varphi /\varphi _{1})^{\alpha }]}~~,
\label{newpot}
\end{equation}
where, $\alpha$ is any odd number, but we adopt to  $\alpha \le 3$ to ensure the renormalizibility.  The parameter $\varphi_1$ then fixes the dip in the potential.\cite{Gangopadhyay15} For appropriate values of the potential parameters, a fit very similar to
that of Figure \ref{fig:1} is attainable.

\subsection{Potential break and cosmic dark flow}
The particular relevance of our work,\cite{Gangopadhyay15} however, is that a break in the inflation-generating potential not only explains the dip, but resolves another problem in early-universe cosmology, namely the magnitude of
cosmic dark flow predicted in the M-theory landscape.\cite{Holman06,Mersini-Houghton09, Kobakhidzr07, Holman09}

The string landscape\cite{Holman06} is a complex space of a huge number of string vacua. These vacua  differ from one another due to the complexity of RR fluxes and whether they are included or not and also the appearance of the gauge groups in the lower energy scale. A full analysis of the structure of the landscape is beyond the scope of present physics.  However,   there should be at least $ 10^{500} $ to $10^{600}$ possible vacua points in the landscape.\cite{Douglas05} Starting from this, a model for the landscape was constructed\cite{Mersini-Houghton09, Kobakhidzr07, Holman09} as a lattice of vacua with a distribution of lattice energies. This landscape lattice-like structure has two sectors SUSY and non-SUSY. Localization disconnects these two sectors, while in the SUSY sector the landscape is viewed as a regular lattice, the non-SUSY sector exhibits an Anderson localization around each vacua site. Accepting that  Supersymmetry is a broken symmetry, then the vacuum energy densities can vary widely being positive, negative or zero.
    
    All landscape vacua have equal probability to give birth to a universe. It has been shown in Refs.~\citen{Mersini-Houghton09, Kobakhidzr07, Holman09} that there are two kinds of baby universes:  one kind will survive the back reaction due to quantum gravitational effects;  and the other  will be terminal universes that will not survive the back reaction. Since all  of the surviving universes were casually connected initially due to  quantum unitarity, there will be an entanglement among all survivors.  This entanglement should be imprinted in our survivor universe as well.     
    
     \subsection{Modified Friedmann Equation}

In the landscape scenario one must  include the back-reaction due to the quantum entanglement with super horizon size wavelength modes.\cite{Mersini-Houghton09, Kobakhidzr07, Holman09}  Then, the modified Friedmann equation becomes:
 
\begin{equation}
 H^{2} = \frac{1}{3M_P^{2}}[V(\varphi)+ \frac{1}{2}(\frac{V(\varphi)}{3M_P^{2}})^{2}F(b,V)]\equiv \frac{V_{eff}}{3M_P^{2}}~~,
\end{equation}
where,
\begin{equation}
 F(b,V)= \frac{3}{2}(2+\frac{m^2M_P^2}{V})log(\frac{b^2M_P^2}{V})-\frac{1}{2}(1+\frac{m^2}{b^2}) \exp{(-3\frac{b^2M_P^2}{V})}~~.
\end{equation}
   Here, $M_P^2$ is $8\pi G_N$, $V(\varphi)$ is the inflation potential, $b$ is the SUSY breaking scale and $m^2=V(\phi)''$. 
   The interference length is given by:
   \begin{equation}
 L_1^2= \frac{a}{H}[(\frac{m^2}{3H})ln\frac{b}{H}-\frac{m^2H}{6}(\frac{1}{b^2}-\frac{1}{H^2})]~~.
\end{equation}

%%%%%%%%%%%%%%%%%%%%%%%%%%%%%%%%%%%%%%%%%%%%%%%%%%%%

    %%%%%%%%%%%%%%%%%%%%%%%%%%%%%%

%%%%%%%%%%%%%%%%%%%%%%%%%%%%%%%%%%%%%%%%%%%%%%%%%%%%%

\subsection{'Tilting' gravitational potential}

The length  scale $L_1$ characterized  the superhorizon size inhomogeneties.  This nonlocal entanglement leads to a modification term $\delta\Phi$ added to the gravitational potential. %%%% tilt gravity potential%%%%%%%%%%%%
\begin{equation} 
 \Phi = \Phi^0 + \delta \Phi\simeq \Phi^0[1+\frac{V(\varphi)F(b,V)}{3M_P^2}(\frac{r}{L_1})]
\end{equation}
where $\Phi^0$ is the inflationary standard gravitational potential and $\delta\Phi$ is the  'tilt'.
This  tilt in the potential is responsible for the dark flow.\cite{Mersini-Houghton09, Kobakhidzr07, Holman09} 

For the quadruple anisotropy we get\cite{Holman09}
%%%%% temperature quad%%%%%%%
\begin{equation}
  \frac{\Delta T}{T} |_{quad}\simeq 0.5(\frac{r_h}{L_1})^2(\frac{V(\varphi)F(b,V)}{18M_P^2})
\end{equation}
Similarly, the effect on anisotropy for dipole is:\cite{Holman09}
 %%%%%% temperature dipole%%%%%
\begin{equation}
 \frac{\Delta T}{T} |_{dip}\simeq \frac{4\pi}{15}(\frac{r_h}{L_1})(\frac{V(\varphi)F(b,V)}{18M_P^2})
\end{equation}

The dark flow velocity (alid for  $L_1\gg r_H$)  is then\cite{Mersini-Houghton09, Kobakhidzr07, Holman09}%$\alpha$ is the calibration factor related to the averaged optical depth of clusters $\tau$
\begin{equation}
  \beta=\frac{v}{c}\simeq \alpha(\frac{\Delta T}{T})|_{dip}\simeq\frac{4\pi}{15}(\frac{r_h}{L_1})(\frac{V(\varphi)F(b,V)}{18M_P^2})
  \end{equation}
   Where we have used 100 km s$^{-1}$ for each $0.4\mu K$ anisotropy in converting the dipole temperature as in Ref.~\citen{Holman09}.
   
    We have revised\cite{Gangopadhyay15} the  theory of Refs.~\cite{Mersini-Houghton09, Kobakhidzr07, Holman09} for a generic inflaton potential and applied it to the specific "break" potential described above in Eq.~(\ref{newpot}).
    % It is is worth mentioning that  the value of optical depth $\alpha$ can be varied.  This  will effect the range of the dark flow potential. So in our work\cite{Gangopadhyay15}  we have considered lower values of $\alpha$. The lower value of $\alpha$ can narrow down the range for the particular form of potential.
Changing the inflaton potential to the new potential in Eq.~(\ref{newpot}) modifies the form of the correction term $F(b,V)$. Since the limits on the SUSY breaking scale are  quite stringent,
 $ 10^{-10}M_P\leq  b \leq  10^{-8}M_P$, the change in the modification mainly comes from the fact the inflaton potential is chaotic-like. This is an interesting case as most of the other possible potentials give much higher  dark-flow velocity.

\begin{figure}[t]
    
  \centering
 \includegraphics[width=0.8\textwidth]{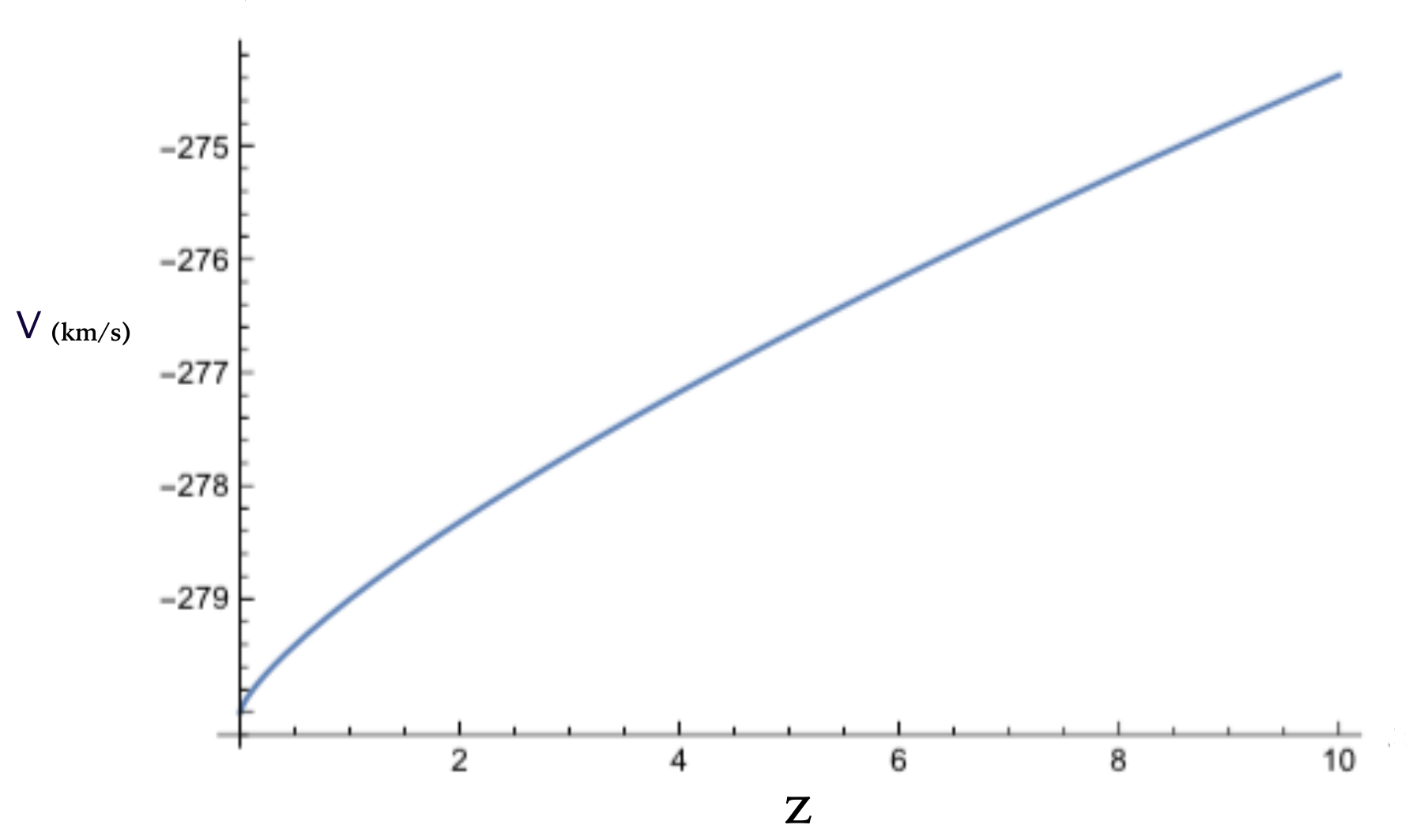}
  \caption{Dark velocity vs z plot with $b\simeq 10^{-10}M_P$}
  \label{figv} 
\end{figure}

        In Figure \ref{figv} from Ref.~\citen{Gangopadhyay15} one can see the dark flow vs.~redshift $z$ for SUSY scale $b\simeq 10^{-10}M_P$. This energy scale can be tested in the distant future in the proposed ILC. If we go to a higher breaking scale we get a higher value for dark flow.
  
  In Refs.~\citen{Mersini-Houghton09, Kobakhidzr07, Holman09} a SUGRA like potential was employed of the form:
  \begin{equation}
 V(\varphi)=V_0\exp{(-\lambda\frac{\varphi}{M_P})}~~.
\end{equation}
Using this potential  a 'dark' flow of  the order of 650-750 km s$^{-1}$ was obtained. In Ref.~\citen{Gangopadhyay15} we have repeated the calculations of Refs.~\citen{Mersini-Houghton09, Kobakhidzr07, Holman09}, but with  with the jump potential of Eq.~(\ref{newpot}). Adopting our  model the resultant dark flow velocity diminishes considerably is illustrated in Figure \ref{figv}.

%%%%%%%%%%%%%%%%%%%%%%%%

  %%%%%%% figure 2%%%%%%%%%%%%
         
%%%%%%%%%%%%%%%%%%%%%%%%%

    %%%%%%%%%%%%%%%%%%%%%%%%%%%%%%
\section{Observational Constraints on Dark Flow}
The above analysis of a lower landscape-generated dark flow velocity is of interest because there are now significant constraints on the bulk flow velocity.  In Mathews et al.\cite{Mathews15b} we have summarized recent constraints on the dark (or bulk) flow velocity, and completed a new analysis based upon an analysis of the SNIa redshift-distance relation.
We made two analyses: one identifying the three dimensional Cartesian velocity components; the other analyzing the cosine dependence on the sky of the deviation from Hubble flow. Fits were for $z < 0.05$ and $z > 0.05$ using both the Union2.1 and SDSS-II supernova surveys. We also studied\cite{Mathews15b} simulated data in which a bulk flow was imposed to determine whether the difficulty in detecting a bulk flow at high redshift is due to uncertainty in the redshift-distance relation, confusion with peculiar velocities, or the absence of a bulk flow. We found\cite{Mathews15b} a bulk flow velocity of $270 \pm 30 \text{ km s}^{-1}$ in the direction $(l,b)=(295 \pm 30, 10 \pm 5)^{\circ}$ in the Cartesian anaylsis, while the cosine analysis gave $325 \pm 50 \text{ km s}^{-1}$ in the direction $(l,b)=(276 \pm 15, 37 \pm 3)^{\circ}$, consistent with previous analyses. In the redshift bin $ z > 0.05$, however, we found\cite{Mathews15b} only marginal evidence for a bulk flow velocity. We also found that the SDSS-II supernova data set has insufficient sky coverage to provide a meaningful result. 

The current observational situation is summarized in Ref.~\citen{Mathews15b} as follows: it
%Searching for a cosmic bulk flow is not new. It
has been known since the 1980's \cite{Lynden-Bell88} that the local dipole flow extends well beyond the local super-cluster.  This source was dubbed ``the Great Attractor."  However, subsequent work in the 1990's \cite{Mathewson92} has shown that the local flow extends at least to $130 ~h^{-1} ~\text{Mpc}$.  However, in Refs. \citen{Darling14,Feix14} peculiar velocity fields were analyzed and found to be  in agreement with a standard $\Lambda$CDM cosmology with no bulk flow, while in other work \cite{Springbob14} an excess bulk flow was deduced. Moreover, there has little evidence of infall back toward  the ``Great Attractor" from material at larger distances.   Although there is recent evidence \cite{Tully14} of a supercluster extending to a scale of $\sim 160~h^{-1}$ Mpc, there remains a need to search for  evidence of a bulk flow at distances well beyond $\sim160 ~h^{-1}~\text{Mpc}$.  Indeed,  a recent analysis\cite{Hoffman15} of the Cosmicflows-2 galaxy peculiar velocity catalog has confirmed  a bulk flow of $250 \pm 21$ km s$^{-1}$ out to a radius of $50~h^{-1}$ Mpc, and inferred  a detectable bulk flow out to  $200~h^{-1}$ Mpc. 
 %New

%old Detecting such dark flow may be possible \cite{Kashlinsky08,Kashlinsky10,Kashlinsky11,Kashlinsky12,Planckdf} by means of the kinetic Sunayev-Zeldovich (kSZ) effect.  This is a distortion
It has been proposed  \cite{Kashlinsky08,Kashlinsky10,Kashlinsky11,Kashlinsky12,Planckdf}  that detecting such dark flow on cosmic scales may be possible by means of the kinetic Sunyaev-Zel\'dovich (kSZ) effect.  This is a  distortion
of the CMB spectrum along the line of sight to a distant galaxy cluster due to the motion of the cluster with respect to the frame of the CMB.  Indeed, a
 detailed analysis \cite{Kashlinsky08,Kashlinsky10,Kashlinsky11,Kashlinsky12} of  the kSZ effect based upon the WMAP data \cite{WMAP9} seemed to indicate the existence of a large bulk flow velocity out to a distance of nearly 800 $h^{-1}$ Mpc.  However, this is an exceedingly difficult analysis \cite{Wright08, Osborne11}, and the existence of a dark flow  has not been confirmed in a follow-up analysis  \cite{Planckdf} using the higher resolution data from the {\it Planck} Surveyor.  The {\it Planck} results set  a (95\% confidence level)  upper limit of $254 \text{ km s}^{-1}$ for the bulk flow velocity and are consistent with no dark flow.   It has, however,  been argued  \cite{Atrio-Barandela13} that the background averaging method in the {\it Planck} Collaboration analysis may have led to an obscuration of the effect.  Indeed, 
a subsequent analysis\cite{Atrio-Barandela14} of the combined WMAP and {\it Planck} data has suggested evidence of a dark flow. %new
Nevertheless, the {\it Planck} Collaboration upper limit is still consistent with as much as half  of the observed CMB dipole corresponding to  a cosmic dark flow.

Independently of whether the dark flow has been detected via the kSZ effect, in view of the potential importance of this effect it is worthwhile to search for such cosmic dark flow by other means. 
 
%Indeed, there have been many attempts \cite{Colin11,Kinney11,Ma11,Weyant11,Turnbull12,Feindt13,Ma13,Rathaus13,Wiltshire13,Feix14,Appleby14} to analyze the redshift-distance relation based upon different standard candles. 
An analysis of peculiar velocity fields is an alternative means  to look for the effect of a dark flow. In particular, an analysis of the distance-velocity relationship (and more importantly its residuals) well beyond the scale of the Local Group is needed to identify a cosmic bulk flow.  A number of attempts along this line have been made 
\cite{Colin11,Dai11,Ma11,Weyant11,Turnbull12,Feindt13,Ma13,Rathaus13,Wiltshire13,Feix14,Appleby14}.

For example, in Refs.~\citen{Ma11,Wiltshire13} the COMPOSITE sample
of $\approx$4500 galaxies was utilized.  %This set includes galaxies  in the nonlinear regime. 
For this set, distances were determined by the Tully-Fisher or  ``great plane" approach, plus a few by SN Ia  distances.  No dark flow was identified for distances greater than about 100 $h^{-1} ~\text{Mpc}$.
%note: %ma11-4536 & Wiltshire13-4534 &  %ma11-150Mpc & Wiltshire13-100Mpc
However, this combined data set might have large systematic uncertainties from the distant objects.  Hence, a more carefully selected data set was subsequently analyzed \cite{Ma13}.  In that analysis the bulk flow magnitude was reduced to $\sim 300 ~\text{km s}^{-1}$ and was only apparent  out to about 80 $h^{-1} ~\text{Mpc}$.

Galaxies in which a SN Ia has occurred provide, perhaps, the best alternative \cite{Colin11,Dai11,Weyant11,Turnbull12,Feindt13,Rathaus13,Appleby14} because their distances are better determined. However there are fewer  data available. There has been a wide variety in the attempts to find a bulk flow in SN Ia data sets. 
%First there have been different data sets used: Union2, 
For example, Dai et al. \cite{Dai11},  utilized the Union2 data set  from the Supernova Cosmology Project\cite{Union2}  and used a Markov Chain Monte Carlo (MCMC) search for the velocity and direction of the bulk flow. Colin et al.\cite{Colin11} did the same but used a different maximum likelihood method. Weyant et al. \cite{Weyant11} utilized  a unique data set as described in their paper. They then used a weighted least squares and a coefficient unbiased method to find the coefficient of the spherical harmonics describing the cosmic expansion. Feindt et al.\cite{Feindt13} analyzed data from the Union2 compilation combined with that from the Nearby Supernova Factory.\cite{Aldering02} They added a coherent motion into their cosmological model to search for a dark flow. 

The issue of sample sparseness has been considered  in a number of works \cite{Weyant11,Rathaus13,Appleby14}. Weyant et al. \cite{Weyant11} found that  biases appear due to a non-uniform distribution or if there is power beyond the maximum multipole in the regression. Rathaus et al. \cite{Rathaus13} also noted effects from large individual errors, poor sky coverage and the low redshift-volume-density, but they were still able to find a statistically significant bulk flow. Appleby et al. \cite{Appleby14} found a bias in galactic latitude and attributed it to the lack of data in the region of $|b|<20^{\circ}$.

Our study\cite{Mathews15b} differed from that of the previous analyses in several key aspects.  We made an independent analysis based upon two different approaches and two separate data sets.  We began with a slight improvement of previous searches\cite{Dai11,Colin11,Feindt13} based upon the Union2 data set  by analyzing $\sim 600$ galaxies with SN Ia redshifts from the Union2.1 supernova data set\cite{Union2.1}.  We utilized  a  MCMC fit to the  three Cartesian components  of bulk flow velocity rather than the velocity magnitude in galactic coordinates.  This approach had better stability near the Galactic pole.  We then also analyzed the same data by searching for a deviation from Hubble flow with a $\cos{\theta}$ angular dependance on the sky.  These studies established the robustness of these two complementary techniques for identifying the magnitude and direction of the bulk flow and confirmed previous detections of a bulk flow out to at least $z = 0.05$.

Having established the viability of the methods adopted\cite{Mathews15b}, we then applied them for the first time to the large sample of ($\sim 1000$) galactic redshifts and SN Ia distances  from the  {\it SDSS}-II survey \cite{Sako07,Frieman08}.  However, we found that the analysis of the {\it SDSS}-II data was severely limited by the paucity of data in the direction of the cosmic dipole moment.  We establish\cite{Mathews15b} that there is a detectable bulk flow at low redshifts, but at best a marginal detection for high redshifts.    

From simulated data sets, we deduced\cite{Mathews15b} that the current uncertainty at high redshifts arises mostly from the current error in the distance modulus. We estimated\cite{Mathews15b} that with a sample like the Union2.1 data set, a detection would require both significant sky coverage of SNIa out to z = 0.3 and a distance modulus error reduction from 0.2 to $\lesssim$0.02 mag. However, a greatly expanded data set of ~3$\times 10^4$ events might detect a bulk flow even with a typical distance modulus error of 0.2 mag asmay be achievable with the next generation of large surveys like LSST.

\section*{Acknowledgments}
Work at the University of Notre Dame is in part supported 
by the U.S. Department of Energy under
Nuclear Theory Grant DE-FG02-95-ER40934 and NASA grant HST-GO-1296901.
Work at NAOJ was supported in part by Grants-in-Aid for Scientific Research of JSPS (26105517, 24340060).
Work at Nagoya University supported by JSPS research grant number 24340048.

%\begin{thebibliography}{000} %for 3 digits
%\begin{thebibliography}{00}  %for 2 digits

\end{document}